\documentclass[aps,prl,reprint,superscriptaddress,showpacs,showkeys]{revtex4-1}
\usepackage{amssymb}
\usepackage{amsmath}
\usepackage{cancel}
\usepackage{color}
\usepackage{braket}
\usepackage{graphicx}
\usepackage{epstopdf}
\usepackage{verbatim}
\usepackage{siunitx}
\usepackage{soul}
\usepackage[toc,page]{appendix}
\usepackage[section]{placeins}

\mathchardef\mhyphen="2D

\begin{document}

\title{All-optical nanoscale thermometry with silicon-vacancy centers in diamond}

\date{\today}

\author{Christian T. Nguyen}
\affiliation{Department of Physics, Harvard University, 17 Oxford St., Cambridge, MA 02138}
\author{Ruffin E. Evans}
\affiliation{Department of Physics, Harvard University, 17 Oxford St., Cambridge, MA 02138}
\author{Alp Sipahigil}
\affiliation{Department of Physics, Harvard University, 17 Oxford St., Cambridge, MA 02138}
\author{Mihir K. Bhaskar}
\affiliation{Department of Physics, Harvard University, 17 Oxford St., Cambridge, MA 02138}
\author{Denis D. Sukachev}
\affiliation{Department of Physics, Harvard University, 17 Oxford St., Cambridge, MA 02138}
\affiliation{P. N. Lebedev Physical Institute of the RAS, Moscow 119991, Russia}
\author{Viatcheslav N. Agafonov}
\affiliation{GREMAN, UMR CNRS-7347, University F. Rabelais, 37200 Tours, France}
\author{Valery A. Davydov}
\affiliation{L.F. Vereshchagin Institute for High Pressure Physics, RAS, Troitsk, Moscow, 108840, Russia}
\author{Liudmila F. Kulikova}
\affiliation{L.F. Vereshchagin Institute for High Pressure Physics, RAS, Troitsk, Moscow, 108840, Russia}
\author{Fedor Jelezko}
\affiliation{Institute for Quantum Optics and Integrated Quantum Science and Technology (IQst), Ulm University, Albert-Einstein-Allee 11, D-89081 Ulm, Germany}
\author{Mikhail D. Lukin}
\thanks{lukin@physics.harvard.edu}
\affiliation{Department of Physics, Harvard University, 17 Oxford St., Cambridge, MA 02138}



\begin{abstract}
We demonstrate an all-optical thermometer based on an ensemble of silicon-vacancy centers (SiVs) in diamond by utilizing a temperature dependent shift of the SiV optical zero-phonon line transition frequency, $\Delta\lambda/\Delta T=$ \SI{6.8(1)}{\giga\hertz\per\kelvin}. Using SiVs in bulk diamond, we achieve \SI{70}{\milli\kelvin} precision at room temperature with a sensitivity of \SI{360}{\milli\kelvin\per\sqrt{\hertz}}. Finally, we use SiVs in \SI{200}{\nano\metre} nanodiamonds as local temperature probes with \SI{521}{\milli\kelvin\per\sqrt{\hertz}} sensitivity. These results open up new possibilities for nanoscale thermometry in biology, chemistry, and physics, paving the way for control of complex nanoscale systems.
\end{abstract}

\maketitle

Nanoscale sensing based on atom-like solid state systems is a new frontier in metrology \cite{Walker:2003uq, Cai:2002kx, Arai:2014fk, Doherty:2012sf, Taylor:2008uq, Balasubramanian:2008kx, Dolde:2011hl, Acosta:2010nr, Kucsko:2013fp, Sildos:2017uq}. Color centers in diamond have attracted particular interest for their robustness to extreme environments and for their ability to be localized within nanometers of the sensing volume. For example, sensing schemes involving the nitrogen-vacancy color center (NV) utilize an electronic spin, microwave, and optical control to measure temperature, magnetic, electric, and strain fields with nanoscale resolution and high sensitivity \cite{Doherty:2012sf, Taylor:2008uq, Balasubramanian:2008kx, Dolde:2011hl, Acosta:2010nr, Kucsko:2013fp}.
\par
A number of other defects have been recently explored using similar control tools for metrology. Recently, the silicon-vacancy color center (SiV) in diamond has emerged as a superior optical emitter with a bright, narrowband zero-phonon line (ZPL) optical transition at room temperature \cite{Hepp:2014qq, Neu:2011rt}. The SiV belongs to a family of interstitial defects whose favorable optical properties arise from inversion symmetry \cite{Goss:2005fk, Iwasaki:2015fk}. The frequency, linewidth and quantum efficiency of the SiV ZPL depend on temperature \cite{Feng:1993fk, Neu:2013uq, Jahnke:2015zl} and can be measured optically, making the SiV a promising platform for thermometry. Here, we demonstrate two approaches based on photoluminescence (PL) and photoluminescence excitation (PLE) spectroscopy to realize all-optical thermometry with SiV ensembles, both in bulk crystal and nanodiamonds. These all optical techniques and the spectral stability of the SiV greatly simplify measurements. This, combined with its emission at relevant wavelengths and the biological compatibility of diamond makes thermometry based on SiVs an attractive candidate for many nanoscale applications \cite{ChemNTherm, ONeal:2004uq}.\\

    \begin{figure}[h!]
		\includegraphics[width=\linewidth]{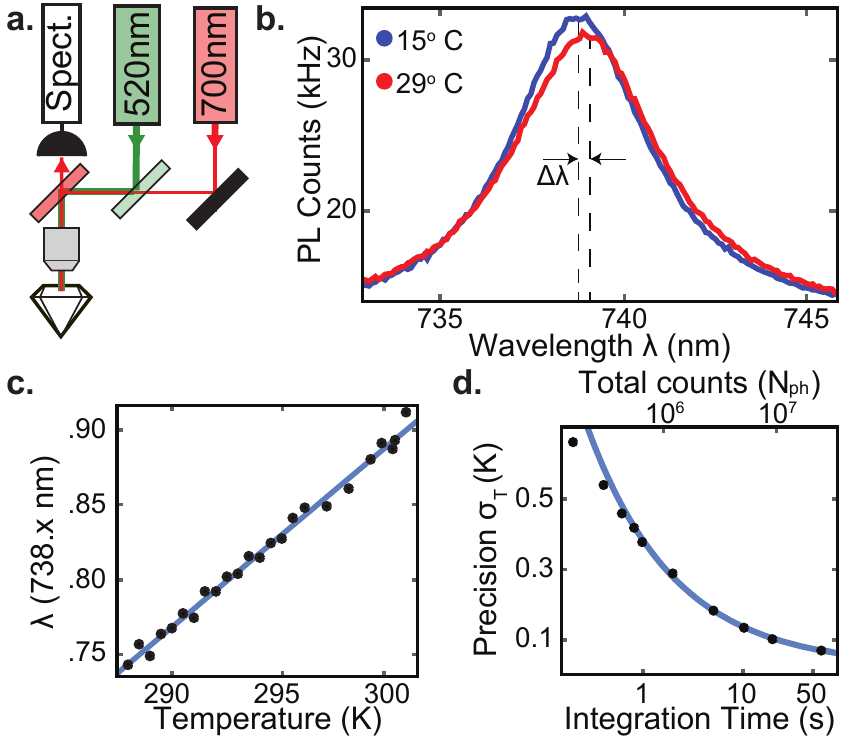}
	\caption{SiV PL thermometry in bulk diamond.
			\textbf{a.} Diamond with a high SiV density is excited off resonance and SiV fluorescence spectrum is measured 
			\textbf{b.} Typical spectra for \SI{15}{\degreeCelsius} (blue) and \SI{29}{\degreeCelsius} (red). The ZPL peak shifts to the red at higher temperatures.
			\textbf{c.} Peak position as a function of temperature. The susceptibility is $\Delta\lambda / \Delta T=0.0124(2)$\si{\nano\metre\per\kelvin} (\SI{6.8(1)}{\giga\hertz\per\kelvin}). Error bars for this measurement are the size of the data points.
			\textbf{d.} Precision ($\sigma_T$) of the thermometer as a function of integration time. Solid line is a fit to shot noise $1/\sqrt{N_{ph}}$. The sensitivity is \SI{360}{\milli\kelvin\per\sqrt{\hertz}}.
			}
	    \label{fig:1}
    \end{figure}
		
\par
Following previous studies \cite{Feng:1993fk, Neu:2013uq, Jahnke:2015zl}, we first focus on the PL spectrum of an ensemble of SiVs in bulk diamond at room temperature. We fit the ZPL spectrum \cite{Foreman-Mackey:2013la, Goodman:MCMC} and use the ZPL peak position as the thermometry signal (PL thermometry). Although the SiV ZPL shifts non-linearly from 5K to room temperature \cite{Feng:1993fk, Neu:2013uq, Jahnke:2015zl}, for a small range ($295\pm 5$\si{\kelvin}), it deviates by less than 1\% from the linear approximation (Fig.\ 1c). We measure the wavelength susceptibility to be $\Delta\lambda / \Delta T=0.0124(2)$\si{\nano\metre\per\kelvin} (\SI{6.8(1)}{\giga\hertz\per\kelvin}) which agrees with previously reported values \cite{Feng:1993fk, Neu:2013uq, Jahnke:2015zl}. The origin of this shift is thermal lattice expansion which reduces the orbital overlap between dangling carbon bonds \cite{Jahnke:2015zl}.
\par
To estimate the precision of SiV thermometry, we measure the uncertainty in the peak position as a function of integration time at fixed temperature (Fig.\ 1d), and extract a sensitivity of \SI{360}{\milli\kelvin\per\sqrt{\hertz}}, giving \SI{70}{\milli\kelvin} temperature precision after \SI{50}{\second} integration time. This measurement uncertainty follows the shot-noise limit $1/\sqrt{N_{ph}}$ (Fig.\ 1d), suggesting that the precision can be improved by increasing photon collection rates from the sample, either by increasing SiV density \cite{Acosta:2009ty, DHaenens-Johansson:2011rz} or by improving the collection efficiency.\\

		    \begin{figure}
		\includegraphics[width=\linewidth]{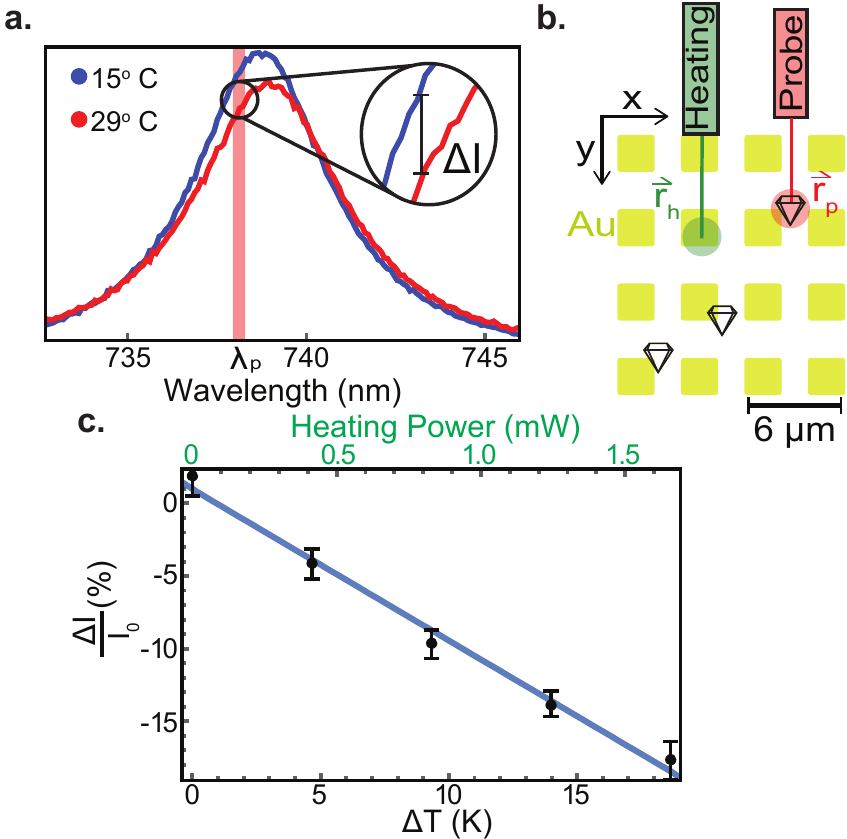}
	\caption{SiV PLE thermometry with nanodiamonds
			\textbf{a.} Scheme for PLE thermometry: Excite SiVs near resonance at $\lambda_p$ and measure PLE fluorescence as a function of temperature ($\Delta I$).
			\textbf{b.} $\sim$\SI{200}{\nano\meter} nanodiamonds are drop cast onto a gold targeting grid. Heat is applied via a green laser at position $\bf{r_h}$, which is probed via the nanodiamond at position $\bf{r_p}$.
			\textbf{c.} Lock-in fluorescence contrast vs.\ temperature for PLE thermometry. The heating laser power $P_h$ is modulated, and PLE fluorescence is measured. The susceptibility is $\frac{\Delta I}{I_0} / \Delta T=1.3(1)$\si{\%\per\kelvin} with sensitivity \SI{521}{\milli\kelvin\per\sqrt{\hertz}}. $\Delta T=T_{P_h\neq 0}-T_{P_h=0}.$
			}
	    \label{fig:2}
    \end{figure}
 \par   
    		
We next demonstrate nanometer-scale thermometry using SiV containing nanodiamonds. For these experiments, we use \SI{200(70)}{\nano\metre} high-pressure high-temperature nanodiamonds grown with silicon contamination in the growth chamber (see methods for growth details). The count rates in these nanodiamonds are a few hundred \si{\kilo\hertz}, suggesting that they contain fewer SiVs ($<10$) in the confocal volume than for the bulk diamond ($\sim100$). With a smaller photon flux, longer integration times are needed in order to achieve the same precision. For our spectrometer CCD, we have a readout noise of $\sim10$ counts per bin and a total of 1500 bins, which limits the detection bandwidth to \SI{0.6}{\hertz} for 10:1 signal to noise (SNR) using the measured count rates, making the measurement sensitive to slow drifts. To overcome this limitation, we introduce a different thermometry technique, PLE thermometry, based on near-resonant excitation of the SiV ZPL transition.
\par
For PLE thermometry, instead of exciting SiVs off resonance and measuring the ZPL spectrum, we excite on resonance (\SI{738}{\nano\metre}  at room temperature), and collect emission into the phonon sideband (PSB), effectively probing the absorption cross-section of the ZPL as a function of temperature. Increasing the temperature red-shifts the ZPL peak and reduces the PLE intensity when the SiVs are excited below saturation, which results in a reduced absorption cross-section blue of the resonance peak. The red-shift is identical to what is measured for PL thermometry, and the intensity reduction arises from non-radiative decays from the excited state becoming more favorable at higher temperatures \cite{Jahnke:2015zl}. These non-radiative decays reduce the quantum efficiency of the ZPL transition and weaken the PLE signal. We therefore excite the nanodiamonds at an experimentally determined wavelength ($\lambda_p$) of maximum contrast (Fig.\ 2a). For this technique, we use an avalanche photodiode (APD) with $\sim50$ dark counts, giving a detection bandwidth of \SI{200}{\hertz} for 10:1 SNR, much larger than for PL thermometry. This high-bandwidth measurement also enables lock-in techniques (described below), which further mitigate slow experimental drifts.
\par
We demonstrate this technique by patterning an array of \SI{2}{\micro\metre} wide \SI{50}{\nano\metre} thick gold pads onto a glass slide using photolithography and drop casting an isopropyl alcohol solution of nanodiamonds containing SiVs (Fig.\ 2b). The gold pads absorb light at \SI{520}{\nano\metre}\cite{Khriachtchev:2001bs}, and act as a local heat source when illuminated. A nanodiamond at position $\bf{r_p}$ (Fig.\ 2b) is continually monitored while the power $P_h$ of a heating laser applied at $\bf{r_h}$ is modulated in a lock-in measurement. This gives rise to a PLE thermometry signal ($\frac{\Delta I}{I_0}$) defined by the normalized difference in counts between $P_h=0$ and $P_h\neq 0$. To calibrate the PLE thermometer, we first measure both $\frac{\Delta I}{I_0} / \Delta P_h$ and $\Delta \lambda / \Delta P_h$. Then, using $\Delta \lambda / \Delta T$ from bulk measurements we extract $\Delta P_h / \Delta T$, and ultimately the susceptibility $\frac{\Delta I}{I_0} / \Delta T = 1.3(1)$\si{\%\per\kelvin}(Fig.\ 2c) with a sensitivity of \SI{521}{\milli\kelvin\per\sqrt{\hertz}}. For nanodiamonds, this deviates from the shot noise limit and is limited to a sensitivity of 1K. This is most likely limited by residual fluorescence noise not rejected by the lock-in technique. Although we should be able to operate at a \SI{200}{\hertz} lock-in modulation frequency based on count rates, the optimal susceptibility occurs around \SI{80}{\hertz}, suggesting the bandwidth for this measurement is limited by the speed of the heating technique itself. While the sensitivity for this technique is worse than that measured using PL thermometry, it is perhaps more fair to compare the two techniques by looking at the sensitivity per sensor volume. For PL thermometry, we have a confocal volume of $\sim$\SI{4}{\micro\meter\cubed} giving a sensitivity per volume of \SI{691}{\milli\kelvin\per\sqrt{\hertz\,\micro\meter\cubed}}. In comparison, nanodiamonds are $\sim$\SI{.004}{\micro\meter\cubed} which gives a sensitivity per volume of \SI{33}{\milli\kelvin\per\sqrt{\hertz\,\micro\meter\cubed}}. For different nanodiamonds on the same substrate, this susceptibility varies by less than $1\%$, which is within error bars for the susceptibility and suggests that the nanodiamonds do not need to be individually calibrated in order to be a precise relative thermometer.

    \begin{figure}
		\includegraphics[width=\linewidth]{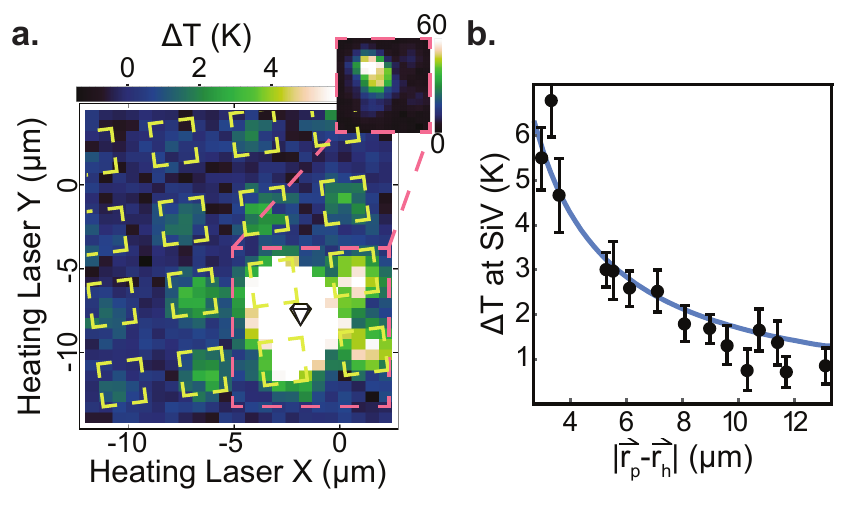}
	\caption{Measuring local heating via PLE thermometry
			\textbf{a.}  Away from the nanodiamond, heating corresponds to the gold array (overlaid boxes). Near the nanodiamond (inset), off-resonant fluorescence from the heating laser dominates the signal. 
			\textbf{b.} Temperature change on gold pads vs. their distance from the nanodiamond. These points follow a $1/|\bf{r_p}-\bf{r_h}|$ dependence as expected from solving the steady-state 2D heat equation\cite{Tadeu:2006kx}, and are consistent with the absorption of a \SI{50}{\nano\metre} gold film at \SI{520}{\nano\metre}\cite{Khriachtchev:2001bs}.
			}
	    \label{fig:3}
    \end{figure}
\par
Finally, we scan $\bf{r_h}$, the position of the heating laser, across our sample to measure the temperature response at the probe position $\bf{r_p}$. Whenever the heating laser passes over a gold pad, the temperature at the nanodiamond increases, leading to the observed pattern in Fig.\ 3a. This map is a measurement of the temperature difference $\Delta T$ at $\bf{r_p}$ induced by the heating laser at $\bf{r_h}$. $\Delta T$ falls off as $1/|\bf{r_p}-\bf{r_h}|$ (Fig.\ 3b), which is consistent with the steady-state solution to the heat equation of a point-source \cite{Tadeu:2006kx}. The fit in Fig.\ 3b has only one free parameter corresponding to the absorption of the gold pads at \SI{520}{\nano\meter}, which is within a factor of 2 of the expected value \cite{Khriachtchev:2001bs}. For such a measurement, the spatial resolution is limited by the nanodiamond size and the precision with which one can focus the heat source. For this experiment, this is the confocal microscope's optical resolution ($\sim$\SI{300}{\nano\metre}).\\

\par

This Article demonstrates nanoscale thermometry based on SiV centers in diamond and achieves a sensitivity of \SI{360}{\milli\kelvin\per\sqrt{\hertz}} in bulk diamond and \SI{521}{\milli\kelvin\per\sqrt{\hertz}} in \SI{200}{\nano\metre}-sized nanodiamonds. Currently, PL thermometry in bulk diamond is shot-noise limited, while PLE thermometry in nanodiamonds is not. This is most likely due to an inability to fully reject fluorescence intensity noise in the lock-in measurement. To address this, one can modify the lock-in technique by modulating two resonant lasers placed on the red and blue sides of the ZPL transition. Since this method does not rely on modulating the sample temperature, this modulation can be done at higher frequencies, which should further reduce noise in the measurement and improve the precision of this thermometry technique. 
\par
Our approach is not only comparable in precision and size to state-of-the-art experiments in nanoscale thermometry \cite{Yue:2012fk, ChemNTherm}, but also possesses several unique benefits. The all-optical scheme eliminates the necessity of microwaves, reducing measurement induced heating and potential damage to the target under investigation. SiVs are embedded in the chemically inert diamond lattice, making them biologically compatible and robust to extreme environments. Additionally, SiVs have a narrow inhomogenous distribution, which makes PL and PLE thermometry with SiVs effective even without individual calibration. Moreover, SiV emission lies in the optical window for \textit{in vivo} imaging, a frequency-band of high transmission for a variety of biological material typically between \SIrange{650}{1350}{\nano\metre} \cite{Smith:2009fk}, rendering SiV thermometry a promising candidate for biological \textit{in vivo} applications \cite{ChemNTherm, ONeal:2004uq}. Finally, recent studies show that nanodiamonds can also be used to vary the temperature of a local environment via optical refrigeration \cite{Kern:2017kx}. This technique would allow SiV incorporated nanodiamonds to be an integrated temperature sensor and actuator at the cellular level.

\section*{Acknowledgements}
We thank J.\ Choi, H.\ Zhou, P.\ Maurer and R.\ Landig for discussions and valuable insight regarding biological applications. Financial support was provided by the NSF, the Center for Ultracold Atoms, the Office of Naval Research MURI, the Gordon and Betty Moore Foundation, and the ARL. FJ was supported by ERC and Volkswagenstiftung. VAD and LFK thank the Russian Foundation for Basic Research (Grant No. 15-03-04490) for financial support.
\section*{Methods}
\subsection{Bulk CVD diamond growth}
In order to achieve a high density of SiVs in a diamond chip, we overgrow a Type IIa diamond using plasma-enhanced chemical vapor deposition (PECVD) in a Seki Technotron AX5010-INT PECVD reactor. A Si wafer is placed in the plasma, where it is etched and incorporated into the overgrown layer. The growth conditions for this layer are microwave power: \SI{950}{\watt}, chamber pressure: \SI{60}{Torr}, gas flow: \SI{300}{sccm} 1:99 $\mathrm{CH}_4:\mathrm{H}_2$ for 20 minutes\cite{Aharonovich:2012fk}. After growth, \SI{1}{\micro\metre} tall islands can be found on the edges of the seed diamond, each of which has a high SiV density.
\subsection{SiV nanodiamond growth}
To grow nanodiamonds, we use high-pressure high-temperature synthesis with luminescent SiV and NV centers based on homogeneous mixtures of naphthalene, fluorinated graphite ($\mathrm{CF}_{1.1}$), and tetrakis(trimethylsilyl)silane without catalyst metals. Cold-pressed pellets of the initial mixture (\SI{5}{\milli\meter} diameter and \SI{3}{\milli\meter} height) are inserted into a graphite container and then placed into a high-pressure cell. High-pressure high-temperature treatment of the samples is performed with the use of a high pressure apparatus of the ``Toroid'' type. Details of the experimental procedure are described in the preliminary report\cite{Davydov:2014ly}. The obtained diamond products are then isolated by quenching to room temperature under pressure.
\par
Separation of nano-size ($\sim$15-400 nm) diamonds was carried out in several stages that consisted of ultrasonic dispersing of the diamond particles using UP200Ht dispersant (Hielscher Ultrasonic Technology), chemical treatment of the samples in a 40\% solution of hydrogen peroxide, and subsequent centrifugation of aqueous or alcohol dispersion of diamond powders. The diamonds used in this work were measured to have a size distribution of \SI{200(70)}{\nano\meter} using the DelsaNano C particle analyzer.

\subsection{PL thermometry setup, fitting procedure, and additional analysis}
In order to measure temperature, the diamond is mounted onto an external heater and excited off resonance with \SI{700}{\nano\metre} and \SI{520}{\nano\metre} light (\SI{10}{\milli\watt} and \SI{5}{\milli\watt} respectively) using a home-built confocal microscope. SiV fluorescence is filtered around the ZPL (Semrock FF01-740/13) and directed to either an APD or a spectrometer (Horiba iHR550 + Synapse CCD, 1800gr, \SI{0.025}{\nano\metre} resolution). Based on fluorescence counts on an APD (\SI{10}{\mega\hertz}), we expect to be addressing $\sim$100 SiVs. Interference effects on the collection arm induce a fluctuating spectral signal modulating our spectrum. To correct for this, we measure the transfer function of a blackbody source traveling through the same optical path and subtract it from the measured signal. As it turns out, fitting a preliminary spectrum to a lorentzian at some fixed temperature and using the fit residuals to find the transfer function agrees to within $10\%$. This method has the added benefit of not requiring a separate calibration step, as the first spectrum in a given measurement can be used for calibration. Afterward, we fit each spectrum to a Lorentzian using an affine-invariant Markov-Chain Monte-Carlo estimator\cite{Foreman-Mackey:2013la, Goodman:MCMC} and extract the ZPL peak position, peak width, and peak intensity, as well as estimated contributions to the noise from both photon shot noise and CCD readout noise\cite{EmceeCode}. While the fit parameters achieved with this technique are almost identical to those acquired using a more naive least-squares fit, this technique is able to give a more convincing estimate for the uncertainties in the fit parameters and agrees with systematic errors extracted via repetitive measurement. As mentioned in the main text, the ZPL peak position shifts by $0.0124(2)$\si{\nano\metre\per\kelvin} with a sensitivity of \SI{337}{\milli\kelvin\per\sqrt{\hertz}}. In comparison, the ZPL linewidth broadens by $0.030(1)$\si{\nano\metre\per\kelvin}  with a sensitivity of \SI{636.7}{\milli\kelvin\per\sqrt{\hertz}}, and the integrated ZPL intensity does not have a statistically significant dependence on temperature. Local strain in the diamond causes different SiV sub-ensembles to have different intrinsic ZPL peak positions and widths corresponding to an offset of $\pm$\SI{2}{\kelvin} at room temperature. Despite this, the susceptibilities vary by only $\pm3\%$, meaning there is no need for calibration in order to measure relative temperature changes. 

\bibliography{TempSensingBib2}
\end{document}